\documentclass{iopart}
\usepackage{iopams} 
\usepackage{times} 

\usepackage{graphicx}
\usepackage{color, bm, cite}
\usepackage{psfrag}

\newcommand{\expt}[1]{\langle #1 \rangle}
\newcommand{\Expt}[1]{\left\langle #1 \right\rangle}

\newcommand{\beq}{\begin{eqnarray}}
\newcommand{\eeq}{\end{eqnarray}}

\newcommand{\pp}{\partial}
\newcommand{\nn}{\nonumber}

\newcommand{\LL}{\mathcal{L}}

\newcommand{\feq}{\rho_{eq}}

\newcommand{\bx}{\mathbf{x}}
\newcommand{\bp}{\mathbf{p}}
\newcommand{\bv}{\mathbf{v}}
\newcommand{\be}{\mathbf{e}}
\newcommand{\bF}{\mathbf{F}}

\begin{document}

\title[Estimating DPD interactions from particle trajectories]{A method for estimating the interactions in dissipative particle dynamics from particle trajectories}
\author{Anders Eriksson$^1$, Martin Nilsson Jacobi$^2$, Johan Nystr\"{o}m$^2$ and Kolbj\o rn Tunstr\o m$^2$}
\address{$^1$ Department of Physics, University of Gothenburg, 41296 G\"oteborg, Sweden} 
\address{$^2$ Complex Systems Group, Department of  Energy and Environment, Chalmers University of Technology, 41296 G\"oteborg, Sweden} 

\begin{abstract}
We introduce a method for determining the functional form of the stochastic and dissipative interactions in a dissipative particle dynamics (DPD) model from projected phase space trajectories. The DPD model is viewed as a coarse graining of a detailed dynamics that displays a clear time scale separation. Based on the Mori-Zwanzig projection operator method we derive a consistency equation for the stochastic interaction in DPD. The consistency equation can be solved by an iterative boot strapping procedure. Combined with standard techniques for estimating the conservative interaction, our method makes it possible to reconstruct all the forces in a coarse grained DPD model. We demonstrate how the method works by recreating the interactions in a DPD model from its phase space trajectory. Furthermore, we discuss how our method can be used in realistic systems with finite time scale separation.
\end{abstract}


\maketitle

\section{Introduction}

The molecular dynamics (MD) method is widely used to predict and analyze systems described by atomic or molecular models. Since only a small volume can be simulated, it is necessary to model how the simulated region interacts with the surroundings to bring the system into equilibrium. A device, or in the context of simulations a procedure, that brings the system to equilibrium is called a thermostat. Depending on the experimental setup the appropriate statistical model of the equilibrated system, the ensemble, is characterized either by constant energy and volume (NVE---MD without thermostat), by constant temperature and volume (NVT--e.g. the Nose-Hoover thermostat \cite{nose84b, hoover85}), or by constant temperature and pressure (NPT--e.g. the Andersen \cite{ andersen80} or Parrinello-Rahman \cite{ parrinello_rahman80} thermostat). Apart from attempting to mimic a specific experimental setup, the choice of thermostat might be guided for example by how easy it is to derive different thermodynamic properties.

We can distinguish properties of the system that depend only on the equilibrium distribution of particle positions and velocities (e.g. the temperature, pressure, radial distribution function and the structure function) from transport properties (e.g. diffusion rate and viscosity). The former quantities depend only on the relative frequency of different states in the distribution, and because the dynamics of molecular systems is usually assumed to be ergodic, thermodynamic properties can be calculated from time averages over a single simulation run, rather than averaged over an explicit ensemble. 

Transport properties, contrary to equilibrium properties, are sensitive to the order in which states occur, i.e., to the temporal correlations. This follows from the Green-Kubo relations \cite{green52,kubo57}, which give the transport properties in terms of autocorrelations of velocities and forces. The thermostat determines how the MD system approaches equilibrium, and its influence on the trajectories in turn alters the autocorrelations (compared to the constant energy dynamics) and therefore the transport properties. In addition to altering transport properties, most thermostats break fundamental symmetries of the systems, such as conservation of linear and angular momentum, since they are typically not Galilean invariant. This leads for instance to a damping of hydrodynamic modes \cite{soddemann_etal03}. In general, any effect of these thermostats on the statistics is therefore an artifact. In order to minimize these effects the strength of the thermostats are typically chosen small enough that the artifacts can be ignored, and the system size is chosen as large as possible.

The effective dynamics of a coarse-grained molecular system has yet another source of dissipative interactions, compared to a system with only atomistic interactions. If the underlying system have a pronounced time scale separation, the thermostat can naturally appear as a result of projecting away the fast degrees of freedom~\cite{mori58,zwanzig60}. MD itself may also be viewed as a coarse-grained representation of underlying quantum mechanics, based on the Born-Oppenheimer approximation. In standard MD the influence of the electron structure is usually neglected (except in \emph{ab initio} MD), and the potentials are fitted to empirical data~\cite{allen_tildesley87}. It is an open question whether a more formal coarse-graining would lead to a natural thermostat for MD systems.

The theoretical framework underpinning this view of the origin of dissipative forces is Mori-Zwanzig theory of projection operators \cite{just_etal03,mori65,mori58,zwanzig_bixon70,zwanzig60}. Briefly, the theory explains how the dynamics of a microscopic system can be mapped to a coarse-grained or mesoscopic level by using projection operators. Depending on the projection, time-scale separation and the degree of exchange of energy between the fast and slow degrees of freedom, the effect of the fast degrees of freedom can either be eliminated due to averaging, resulting in a deterministic dynamics for the slow degrees of freedom, or it can result in Markovian white noise and dissipation, leading instead to a stochastic dynamics described by a Fokker-Planck equation \cite{zwanzig02}. Used in equivalent Langevin-type equations, the drift and diffusion term in the Fokker-Planck equation naturally defines the thermostat for the coarse-grained system.

An example of Langevin-type dynamics that can be coupled to the Mori-Zwanzig theory is the simulation technique dissipative particle dynamics (DPD). DPD is a particle based coarse grained model with pairwise central force interactions. The interactions have both a conservative part and a part given by noise and dissipation. By construction, the DPD model is Galilean invariant and can therefore be used to simulate nontrivial hydrodynamics. It was first suggested as a simulation tool for complex fluids\cite{hoogerbrugge_koelman92}, using soft conservative potentials. This has been the major conception of DPD, but it has also been shown that the method is suitable as an alternative thermostat for MD simulations \cite{soddemann_etal03}.

An important conceptual shift is introduced if the dissipative and stochastic forces in DPD do not stem from interactions with the surroundings of the system but from interactions between the coarse-grained degrees of freedom and the degrees of freedom not explicitly modelled. In this situation, the thermostat should no longer be viewed as merely a means to make the system approach equilibrium,  but as an integral part of the dynamics, where both the functional form and the strength of the thermostat are defined by the projection from the microscopic to the mesoscopic system. There exists formal coarse graining schemes resulting in DPD like mesoscopic dynamics\cite{espanol98, flekkoy_etal99}.  The main idea in this paper is to introduce a practical method for determining the functional form of the stochastic interactions in the DPD model from the phase space trajectories of the coarse grained system. We apply  Mori-Zwanzig theory to derive a consistency expression that the stochastic interaction in the DPD model must fulfill. This result is used to derive a practical boot strapping method that can be used with simulation data to obtain a realistic estimate of the full functional forms of the effective coarse-grained interactions. In order to demonstrate the method and test its consistency, we apply it to phase space trajectories from a DPD simulation with known conservative, dissipative and stochastic forces.

\section{Theoretical analysis}

This section describes how the DPD model can be viewed as the effective dynamics resulting from a projection of an underlying atomistic dynamics. We begin with a general review of the projection operator method. We then discuss how DPD can be used as a specific ansatz for the effective dynamics resulting from center of mass projections of atomistic systems. We discuss the equilibrium and transport properties separately: First, for given stochastic forces, we show how the dissipative force must be chosen to maintain the equilibrium distribution, and how this relation leads to dissipative forces that respect the fundamental symmetries of the underlying dynamics. Second, we show that the combined effect of the stochastic and dissipative forces is to drive the system to thermal equilibrium, and that the radial dependence of the stochastic forces determines the rate of convergence to thermal equilibrium.
In summary, this shows that the dissipative and stochastic parts of the effective coarse-grained dynamics are not arbitrary but are determined by the underlying dynamics through the choice of projection.

\subsection{Projection operators}\label{sec:projection}

Consider a dynamical system 

\begin{eqnarray}
\dot{x} & = & f(x,y), \\
		\dot{y} & = & \epsilon ^{-1} g(x,y) ,
		\end{eqnarray}
consisting of fast ($y$) and slow ($x$) degrees of freedom, and a time scale separation indicated by the parameter $\epsilon \ll 1$. The corresponding Liouville operator splits into a fast and a slow part:
\begin{equation}
	-  \mathcal{L} = \underbrace{\sum _i \frac{\partial}{\partial x_i} f_i (x , y)}_\textrm{\small slow} + \underbrace{\frac{1}{\epsilon} \sum _i \frac{\partial}{\partial y_i} g _i (x,y)}_\textrm{\small fast}  .
\end{equation}
From the time evolution in density space, $\partial _t \rho = - {\mathcal L} \rho$, an effective Fokker-Planck equation for the slow degrees of freedom can be derived. Following Just et al.~\cite{just_etal03}, consider the adiabatic distribution $\rho _{ad} (y |x)$ as the stationary distribution of $y$ when $x$ is considered fixed (changing adiabatically slowly). Under the assumption of ergodicity, an adiabatic average  over the fast degrees of freedom conditioned on the slow degrees of freedom can be defined for an arbitrary function $h$,

\begin{equation}
	\langle h \rangle _{\! ad} (x)  =   \int \!\! d y \, h(x,y) \rho _{ad} ( y | x ) = \lim _{T \rightarrow \infty} \frac{1}{T} \int _0 ^T \!\!\! d \tau \, h ( x , \eta [ \tau , y  ; x ] ) ,
\label{adiabatic_proj}
\end{equation}
where   $\eta [ \tau , y  ; x ]$ is the trajectory of the differential equation $\dot{\eta} = \epsilon ^{-1} g(x,\eta)$ with $\eta (0) = y$ and $x$ fixed. The ergodicity relation can be used in a practical situation to derive an adiabatic average from the detailed trajectory. The reduced phase space density can now be defined as $\bar{\rho}= \int dy \rho _{ad} (y | x)$, and the corresponding Fokker-Planck equation takes the form 

\begin{equation}
	\frac{\partial \bar{\rho} _t}{\partial t} = - \sum _i \frac{\partial}{\partial x _i} D_i ^{(1)} (x) \bar{\rho} _t (x) + \sum _{i j} \frac{\partial ^2}{\partial x_i \partial x_j} D _{i j} ^{(2)} (x) \bar{\rho} _t (x) ,
\label{FP_final}
\end{equation}
where the diffusion term is defined as
\begin{equation}
	D_{i j} ^{(2)} (x) =  \int _0 ^{\infty} \!\!\!\!\! \rmd \tau \, \big< \delta _F \!  f_i (x,  \eta [ \tau , y ; x ] ) \, \delta _F \! f_j (x,   y) \big> _{\! \! ad},
\label{MZ_diffusion}
	\end{equation}
and we use the notation
\begin{equation}\label{eq:deltaf}
	\delta _ F \! f_i (x,y) = f _i (x,  y) - \langle f _i \rangle _{\! ad}  (x)
\end{equation}
as an abbreviation  for the fluctuations around the adiabatic equilibrium. The diffusion coefficient is given by the auto-correlation of the fluctuations in the fast degrees of freedom around their adiabatic stationary mean value. This is the relation that we will use to derive the functional form of the noise in DPD. At equilibrium, the drift term $D^{(1)}$ is derived from the diffusion term using the fluctuation-dissipation theorem. 

In the form presented here, the Fokker-Planck equation represents the global dynamics on the phase space. There is no assumption that the system should have the structure of a mechanical system consisting of particles with pairwise interactions. For the methodology to be useful it is necessary to adopt it to a situation where an effective particle dynamics can be derived, e.g. a model with pairwise additive interactions like the DPD model. In Section~\ref{sec:estimate} we continue the discussion on projection operators by showing how to apply the theory to derive a DPD dynamics from a more detailed, typically deterministic, simulation of a many-particle system.

\subsection{The DPD model}\label{sec:dpdmodel}

In a many-particle simulation a natural choice of coarse graining is to  group  particles in an underlying description of the system, e.g. an atomistic model, into single spherical particles (beads) positioned at the center of mass of the underlying particles. Furthermore, since forces between particles in the underlying system typically are pairwise and opposite, so that the system obeys Newton's third law, we assume that this also holds for the effective forces on the coarse-grained particles. This guarantees that the coarse-graining procedure does not break the conservation of linear and angular momentum, which in turn guarantees that the proper hydrodynamical behavior will be preserved in the approximation. The forces can be divided into three categories: conservative and dissipative deterministic forces, and stochastic forces. The conservative forces stem directly from the conservative interactions between the microscopic particles in one bead with the particles in another bead. The stochastic forces is the result of how fast chaotic degrees of freedom of the particles in each bead fluctuate around the motion of the center of mass and give rise to rapidly fluctuating forces. Finally, the dissipative forces represent the combined effect of the fast degrees of freedom on the slow degrees of freedom. With particles positioned at $\mathbf{r}_{i}$, with velocities  $\mathbf{v}_{i}$ and momenta $\mathbf{p}_{i}$, the equations of motion for a DPD model can be written as a system of Langevin equations
\begin{eqnarray}\label{eq:simple_langevin}
	\dot{ \mathbf{r} }_{i}    &= \mathbf{v}_{i}, \nonumber \\
	\dot{ \mathbf{ p } }_{i}  &= \sum_{j \neq i} \left[ \mathbf{F}_{ij}^{C} + \mathbf{F}_{ij}^{D}
										  +\mathbf{F}_{ij}^{S} \right],
\end{eqnarray}
where $\mathbf{F}_{ij}^{C}$, $\mathbf{F}_{ij}^{D}$ and  $\mathbf{F}_{ij}^{S}$ are the conservative, dissipative, and stochastic forces between particles $i$ and $j$. In DPD, the stochastic force between particles $i$ and $j$ take the form
\beq\label{eq:f_stoch}
	\mathbf{F}_{ij}^{S} =  \sqrt{2k_{B}T}\omega( r_{ij} ) \, \zeta_{ij} \, \mathbf{e}_{ij},
\eeq
where $r_{ij}$ is the distance between particles $i$ and $j$, $\mathbf{e}_{ij} $ is the unit vector pointing from $j$ to $i$, $k_{B}$ is Boltzmann's constant, and $T$ is the temperature in Kelvin. The scalar function $\omega(r_{ij})$ describes how the stochastic force depends on the distance between the particles, and $\zeta_{ij}$ is interpreted as a symmetric Gaussian white noise term with mean zero and covariance
\begin{equation}\label{eq:zeta_cov}
	\expt{\zeta_{ij}(t)\zeta_{i'j'}(t')} = (\delta_{ii'}\delta_{jj'} + \delta_{ij'}\delta_{ji'})\delta(t-t'), 
\end{equation}
where $\delta_{ij}$ and $\delta(t)$ are the Kronecker and Dirac delta functions, respectively. This structure of the covariance matrix makes sure that the stochastic forces between any pair of beads are central forces with equal magnitude, thus preserving the linear and angular momentum of the system.

In order to illustrate how the DPD dynamics acts as a thermostat, and to further emphasize that the dissipative and stochastic parts of the dynamics are not arbitrary but are determined by the underlying dynamics through the choice of projection, we first derive how the damping force must be chosen in order to maintain the proper thermal equilibrium distribution of velocities and positions for a general Hamiltonian system, and then show how the radial dependence of the stochastic forces determine the rate of convergence to the equilibrium distribution, starting from an arbitrary distribution over phase space. 

\subsection{Equilibrium dynamics}\label{sec:equidyn}

At thermal equilibrium the system is distributed according to the canonical ensemble
\begin{equation}\label{eq:feq}
   \feq(\bx,\bp) = Z^{-1} \rme^{-H(\bx,\bp)/k_{B} T},
\end{equation}
where $Z$ is the normalization term for the distribution. Since the equilibrium distribution depends only on the temperature and the Hamiltonian of the system, the conservative forces uniquely determine the equilibrium distribution. The converse is also true; when the conservative forces depend only on the distance between particles, as in the DPD dynamics, it is possible to use an iterative approach to uniquely determine the conservative forces from  the equilibrium radial distribution \cite{lyubartsev1995, soper96, lyubartsev03, reith_etal03, almarza_lomba03, ilpo04} (alternatively, direct time averaging over the fast degrees of freedom can be used, see e.g. Refs. \cite{forrest_suter95,izvekov_parrinello04}). The methods for reconstructing effective potentials from the RDF rely on the result by Henderson\cite{henderson74}, that two pairwise potentials resulting in the same RDF cannot differ by more than an additive constant. The importance of this theorem lies in the one-to-one correspondence between pairwise central force and the radial distribution function.

The question is now: What is required of the forces to maintain the equilibrium distribution? The equilibrium ensemble is automatically invariant under the conservative parts of the dynamics, since $H$ is a constant of motion for Hamiltonian dynamics (this is true for any ensemble where the probability of finding the system in a given micro-state depends only on the energy). 
The stochastic and dissipative forces generally change the equilibrium distribution, except if we choose the dissipative force $\bF^{D}_i$ such that the combined contributions from the dissipative and stochastic forces cancel when acting on the equilibrium distribution (the fluctuation-dissipation relation). Writing down the Fokker-Plank equation that describes the time evolution of the distribution over the state space in the DPD equations of motion, and requiring that the equilibrium distribution is a stationary point of the dynamics, leads to \cite{eriksson_2008b}
\beq\label{eq:equil_cond}
   0 &= \LL \rme^{-H(\bx,\bp)/k_{B}T} \nn\\
	  &= \sum_i \nabla_{\bp_i} \!\cdot\! \Big[ - \mathbf{F}^{D}_i + \frac{1}{2} \sum_j 2 k_{B}T A_{ij}(\bx) \nabla_{\bp_j} \Big]  \rme^{-H(\bx,\bp)/k_{B}T} \nn\\
	  &= 	\sum_i \nabla_{\bp_i} \!\cdot\! \Big[ -  \mathbf{F}^{D}_i  - \sum_j A_{ij}(\bx) \nabla_{\bp_j} H \Big]  \rme^{- H(\bx,\bp)/k_{B}T}
\eeq
where $\LL$ is the Fokker-Planck operator of Equation (\ref{eq:simple_langevin}), and $A_{ij}$ is a $3 \times 3$ matrix given by the covariance of the total forces on particles $i$ and $j$.  The equilibrium Fokker-Planck equation (\ref{eq:equil_cond}) is commonly referred to as a fluctuation--dissipation relation.  Since it must hold for all points $(\bx,\bp)$ in phase-space, the only possible solution for the dissipative force is 
\begin{equation}\label{eq:F_d_sol}
	\mathbf{F}^{D}_i = - \sum_j A_{ij}(\bx) \nabla_{\bp_j} H.
\end{equation}

We briefly comment on the role of the coarse-graining projection in determining the dissipative forces.
In DPD it is usually assumed that the projection is from a set of underlying particles to their center of mass.
If this is not the case, but a more general projection is used, the equilibrium distribution of the projected dynamics is not necessarily the Gibbs distribution with a standard quadratic kinetic term. Especially, it may mean that the momentum-dependent part of the distribution depends also on space (i.e. that it is not possible to split the distribution into a momentum term $\exp(-\sum_i \bp_i^2/2m_i k_B T)$ and a term that depends only on the positions of the DPD particles). This, in turn, means that the damping force may have a different shape than it has now (especially it may not be simply linear in the velocity). However, Equation~(\ref{eq:F_d_sol}) is still valid as long as there exist an energy function $H$ such that the Gibbs distribution describes the equilibrium.

For the DPD model the force covariance is given by
\begin{equation}\label{eq:corr_mat}
A_{ij}  = \cases{ 
      -\omega^2(r_{ij})\, \be_{ij} \otimes \be_{ij} & when $i \ne j$ \\
		\sum_{k\ne i} \omega^2(r_{ik})\, \be_{ik} \otimes \be_{ik} & when $i = j$,}
\end{equation}
where $\otimes$ denotes an outer product, and $\omega(r)$ determines the radial dependence of the stochastic force.
Inserting Equation~(\ref{eq:corr_mat}) into Equation~(\ref{eq:F_d_sol}) we can write the dissipative force on a particle as a sum of pair-wise dissipative forces:
\begin{equation}\label{eq:F_d_sol2}
	\mathbf{F}^{D}_i = \sum_{j \ne i} \mathbf{F}^{D}_{ij} 
	                 = - \sum_{j \ne i} \omega^2(r_{ij})\, \be_{ij} \cdot (\bv_i - \bv_j) \, \be_{ij}.
\end{equation}
Equation~(\ref{eq:F_d_sol2}) was first derived by \cite{espanol_warren95}.
Note that since $\mathbf{F}^{D}_i$ depends only on the velocity differences between interacting particles, it is manifestly Galilean invariant, and it is clear from the derivation above that this is a direct consequence of the covariance property of the stochastic forces [c.f. Equation~(\ref{eq:zeta_cov})], which in turn stems from the assumption that Newton's third law applies.

\subsection{Global convergence to equilibrium}

We have seen that the conservative part of the dynamics is determined by the equilibrium distribution, and the dissipative part of the dynamics is determined by the dependence of the Hamiltonian on the momentum of the beads in combination with the structure of the stochastic forces. 
This leaves only the stochastic forces to be determined in order to have a complete description of the DPD dynamics. The equilibrium distribution gives no hint here, since for any choice of stochastic force the dissipative force will maintain the Gibbs distribution. Instead, the choice of stochastic force will determine how the system approaches equilibrium.
In order to better understand the effect of the stochastic forces on the path to thermal equilibrium, it is illuminating to study the time-evolution of the Kullback distance \cite{kullback59} from the non-equilibrium ensemble $\rho(\bx,\bp)$ to the equilibrium distribution $\feq(\bx,\bp)$, given by 
\begin{equation}\label{eq:entropy}
  K(t) = \int\rmd \bx\,\rmd \bp\, \rho(\bx,\bp) \, \ln \frac{\rho(\bx,\bp)}{\feq(\bx,\bp)}.
\end{equation}
The Kullback distance is non-negative for all ensemble distributions, and is zero if and only if the distribution is identical to the equilibrium distribution. With strictly Hamiltonian dynamics, $K(t)$ is constant in time. Intuitively, this is because the internal energy of the system needs to change in order for the ensemble to approach the equilibrium distribution, but the Hamiltonian conserves the energy. Using the full Fokker-Planck equation of the DPD dynamics, we can calculate the rate of change of $K(t)$:
\beq\label{eq:entropy-change}
 \pp_t K(t) 
  &= - k_{B} T \int\rmd \bx\,\rmd \bp\, \rho \sum_{ij} \left( \nabla_{\bp_i} \log \frac{\rho}{\feq} \right)^T A_{ij} 
                                                       \left( \nabla_{\bp_j} \log \frac{\rho}{\feq} \right) \nn\\
 &= - \frac{k_{B} T}{2} \int\rmd \bx\,\rmd \bp\, \rho 
  		\sum_{i\ne j} 
		\biggl[
      \omega(r_{ij}) \, \be_{ij}  \!\cdot\!
      \bigl(\nabla_{\bp_i} - \nabla_{\bp_j}\bigr)\log \frac{\rho}{\feq}
		 \biggr]^2
\eeq
which is negative, except when the system is at equilibrium. Note that the change in $K(t)$ does not depend directly on the conservative forces, but on the structure of the stochastic forces, $\omega(r)$. For any choice of $\omega(r)$, and from any initial distribution $\rho$ over the phase space for the particle system, $K(t)$ continues to decrease until $\rho = \feq$, where $K = 0$.

Consider an initial distribution $\rho(\bx,\bp)$ such that the system is in equilibrium with respect to the momentum space, but not with respect to position space (i.e. the momentum dependence of $\rho$ and $\feq$ is the same, but the position dependence is different). In this case, we see that $\pp_t K = 0$ despite the system being out of equilibrium. This apparent paradox is resolved by calculating to the second derivative of $K$,  to see that $K$ is still a decreasing function of time ($\pp_t^2 K < 0$); such distributions correspond to saddle points in the dynamics.

From this derivation it is apparent that the dissipative and stochastic forces act as a thermostat to bring the system to the equilibrium distribution, and that the rate at which this occurs, and by which path it occurs, is determined by the structure of the stochastic force (in combination with the conservative force). Thus, if we want the transport properties of our DPD system to match those of the underlying system, it is necessary to find the correct choice of stochastic force. In the next section, we describe a practical method for estimating the stochastic force function $\omega(r)$ from observed trajectories of the system.

\section{Estimating the stochastic force}\label{sec:estimate}

Coarse grained models of molecular systems are assumed to represent the projected dynamics of an underlying more detailed model. In the cases when the detailed dynamics can be simulated e.g. by the MD method, example trajectories of the projected system can be extracted by applying an explicit projection to the trajectories from the MD system. The detailed simulations of smaller systems over shorter time scales can then be used to calibrate, or as we do in this paper derive, the effective interactions in  the coarse grained model. In this section we show that if the resulting reduced model fulfill the DPD ansatz (as formulated in \Sref{sec:dpdmodel}), the observed trajectories contain enough information to estimate both the deterministic and stochastic forces in the DPD equations. As mentioned in \Sref{sec:equidyn} it is well established how the conservative force is determined from the equilibrium radial distribution function. In this section we propose a method for estimating $\omega(r)$, determining both the dissipative and stochastic forces.

\subsection{Relation between DPD and the Mori-Zwanzig therory}

In order to relate the stochastic force in the DPD equations of motion to the observed behaviour of the particles, consider a pair of beads, $i$ and $j$, a distance $r$ from each other at time zero. As a first approach (and a naive one as we will explain), we expand the particle positions and momenta at a short time $\tau$ to obtain
\beq\label{eq:dpidpj}
	\Expt{\delta_t \bp_i \cdot \delta_t \bp_j}_r|_{\tau } = - 2 k_B T \omega^2(r) \tau  + \Or(\tau ^2)
\eeq
where $\Expt{}_{r}$ denotes a conditional ensemble average over all conformations where $r_{ij} = r$, and $\delta_t \mathbf{p} = \mathbf{p}(\tau ) - \mathbf{p}(0)$. For small $\tau $, only the leading term will be important, and can be used to estimate $\omega^2(r)$. 

\Eref{eq:dpidpj} relies explicitly on the infinite time-scale separation between the stochastic and deterministic forces. Such a separation can never hold in a natural system, since the coarse grained dynamics is smooth on the time scale of the underlying dynamics;
this means that in the time region where \Eref{eq:dpidpj} can be used to estimate $\omega(r)$, the DPD ansatz is not valid.

A better approach follows from the Mori-Zwanzig coarse graining scheme laid out in \Sref{sec:projection}, where the structure of the stochastic force was formulated in terms of a Green-Kubo relation \eref{MZ_diffusion}. To make contact between the general Mori-Zwanzig theory and the DPD model, the global dynamics is split into pairwise interactions, and the relative distances between pairs of particles and their relative velocities are assumed to be slow variables. Then, by comparing the stochastic term in the DPD model with \eref{MZ_diffusion}  we find
\beq\label{eq:omegaSq}
	\omega^2(r) = -\frac{1}{k_B T}\int_{0}^{\infty} \rmd t \Expt{ \delta_F \mathbf{F}_i(t) \cdot \delta_F \mathbf{F}_j(0)}_r,
	\label{omegaSq}
\eeq
where $\delta_F \mathbf{F}_i$ is the difference between the projected total force on the beads and the adiabatic average, i.e., \Eref{eq:deltaf}
 adapted to the DPD ansatz. It is important to note that the adiabatic average is defined by all the deterministic forces on the reduced level, both conservative and dissipative. 
 
Rather than attempting to calculate the integral in \Eref{eq:omegaSq} directly, we want to show how this can be estimated from observed trajectories of the coarse grained system. Using the notation
\beq\label{eq:deltadelta}
  \delta_t (\delta_F \bp_i) = \int_0^{\tau } \rmd t\, \delta_F \mathbf{F}_i(t), 
\eeq
we find the identity
\beq
	2\int_{0}^{\tau } \rmd t \Expt{ \delta_F  \mathbf{F}_i(t) \cdot \delta_F  \mathbf{F}_j(0)}_{r}
	= \frac{\partial}{\partial \tau } \Expt{\delta_t (\delta_F \bp_i)\cdot\delta_t (\delta_F \bp_j)}_{r} .
\eeq
Thus, we can express $\omega^2(r)$ in terms of the asymptotic slope of the momentum change covariance:
\beq\label{eq: final derived omega}
	\omega^2(r) = \lim_{\tau \rightarrow \infty}-\frac{1}{2 k_B T} \frac{\partial}{\partial \tau} \Expt{\delta_t (\delta_F \bp_i)\cdot\delta_t (\delta_F \bp_j)}_{r} .
\eeq
In practice it is not possible to take the limit to infinity, due to the so-called plateau problem \cite{espanol_zuniga93}. This arises because for large values of $\tau $, the slope of $\Expt{\delta_t (\delta_F \bp_i)\cdot\delta_t (\delta_F \bp_j)}_{r}$ must necessarily vanish, since eventually all beads separate and move independently. In addition, the measurements are conditioned on a distance $r$, rendering the right side of \Eref{eq: final derived omega} meaningless in the limit of large $\tau$ since the constant $r$ cannot be defined.
If the fast and the slow parts of the dynamics are sufficiently well separated, however, there is a region where $\tau $ is large enough that the fast degrees are in an approximate local (thermal) equilibrium determined by the slow degrees of freedom, but small enough that the slow degrees of freedom do not have time to change significantly. In this region, $\Expt{\delta_t (\delta_F \bp_i)\cdot\delta_t (\delta_F \bp_j)}_{r}$ is approximately linear, and the slope can be used to estimate the stochastic forces. If the conditioning on $r$ does not hold exact, this can result in a small perturbation of $\Expt{\delta_t (\delta_F \bp_i)\cdot\delta_t (\delta_F \bp_j)}_{r}$. How to deal with this in practice will be explained in the next section.
  
\subsection{Boot strapping method}\label{estimate}

The operator $\delta _F$ introduces a dependence of the dissipative force for the right side of \Eref{eq: final derived omega}. The dissipative force in turn depends on $\omega (r)$ (see \Eref{eq:F_d_sol2}) and \Eref{eq: final derived omega} is therefore not closed. This can be resolved by a ``boot strapping'' approach.

To estimate $\omega(r)$ we assume access to a set of coarse grained trajectories, with the same time resolution as the underlying dynamics. $\omega(r)$ is found by solving \Eref{eq: final derived omega} iteratively, with e.g. $\omega(r)=0$ in the first iteration. In the calculations, $\tau$ is typically chosen to be much larger than the time steps in the underlying dynamics. 

\begin{figure} 
\centering 
\psfrag{xlabel}[][]{$\tau$}
\psfrag{ylabel}[][]{$-\Expt{\delta_t (\delta_F \bp_i)\cdot\delta_t (\delta_F \bp_j)}_{r}$}
\psfrag{A}[][]{$r_1$}
\psfrag{B}[][]{$r_2$}
\psfrag{C}[][]{$r_3$}
\includegraphics[width=8cm]{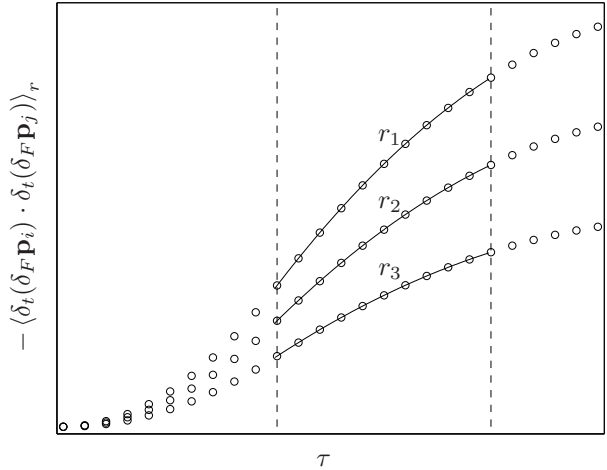}
\caption{\label{fig: minus_DdpiDpdj_FC25}
The figure shows a sketch of typical measurements of the momentum change covariance. Each curve (circles) corresponds to a given value of $r$ and for clarity only a few values of $r$ are shown. In the left section, the dynamics of the coarse grained system is on the same time scale as the underlying dynamics and is therefore smooth. The mid section shows the region where the DPD ansatz is supposed to hold. This region should be approximately linear, but can have higher order terms, as explained in the text. A second order polynomial fit is used in this region (solid lines). An estimate of $\omega^2(r)$ is obtained from the coefficients of the linear terms.}
\end{figure}

The iteration procedure starts with the calculation of $\Expt{\delta_t (\delta_F \bp_i)\cdot\delta_t (\delta_F \bp_j)}_{r}$ as a function of $\tau$ for each value of $r$. To obtain an estimate of $\omega(r)$, the time region where the DPD ansatz is expected to be valid must be identified. This can be done by visual inspection, as illustrated in \Fref{fig: minus_DdpiDpdj_FC25}: For small values of $\tau$ (left section in \Fref{fig: minus_DdpiDpdj_FC25}), the coarse grained dynamics follows the underlying dynamics smoothly and cannot be expressed in terms of DPD. Thereafter, the time region of interest follows (mid section in \Fref{fig: minus_DdpiDpdj_FC25}). This region should be approximately linear in $\tau$, but this might not hold for two reasons. First, the measurements are conditioned on $r$ being constant, but $r$ is actually changing (slowly) with $\tau$. Second, unless the boot strapping procedure has converged, the dissipative force is not correct.  Both these factors will affect the shape of the curves in  \Fref{fig: minus_DdpiDpdj_FC25}. To compensate for this, we fit the curves to a second order polynomial in the selected time region. Under the assumption that the second order terms are mainly a result of the conditioning on $r$ (and the dissipative force when this has not yet converged), the coefficients of the linear terms give for each value of $r$ the estimate of $\omega^2(r)$. This procedure is repeated until convergence. As a word of caution, we recommend not to calculate the direct numerical derivative of the curves, as \Eref{eq: final derived omega} suggests. The numerical differentiation introduces noise and requires significantly longer simulations to obtain good statistics. It is much better to first do a (local) fit of the curve to a low-order polynomial and then evaluate the derivative of the fitted curve \cite{pre96:num}.
	
\section{Numerical verification}\label{sec: numerical}

In order to verify the applicability of the proposed boot strapping method, we test the method on a DPD system with known $\omega(r)$. The DPD simulations were set up using the standard implementation from Groot and Warren~\cite{groot_warren97}, with a density of $4.0$ and a time step of $0.005$. This time step is rather small, but we assume here that for a real coarse grained system, we will have access to the microscopic dynamics, which evolves on a time scale smaller than that usually used in DPD. 
The conservative force was chosen as
\begin{equation}
\label{eq: linear DPD force}
{\bf F}^C_{ij} = \cases{
		a\left(1-r_{ij}\right)\hat{\bf r}_{ij} & when $r_{ij} < 1$, \\
		0 & otherwise,
	}
\end{equation}
with $a = 25$, and $\omega(r)$ was chosen as,
\begin{equation}
\label{eq: linear omega function}
\omega(r_{ij}) = \cases{
		\sigma\left(1-r_{ij}\right) & when  $r_{ij} < 1$, \\
		0 & otherwise,
	}
\end{equation}
with $\sigma = 3.0$. These are standard parameter values chosen for the DPD fluid to match the compressibility of water at room temperature \cite{groot_warren97} and has been used extensively in mesoscopic simulations of lipid membranes \cite{shillcock_lipowsky02,yamamoto_etal02}.

\begin{figure} 
\centering
\psfrag{r}[t][]{$r$}
\psfrag{omega}[b][]{$\omega(r)$}
\includegraphics[width=8cm]{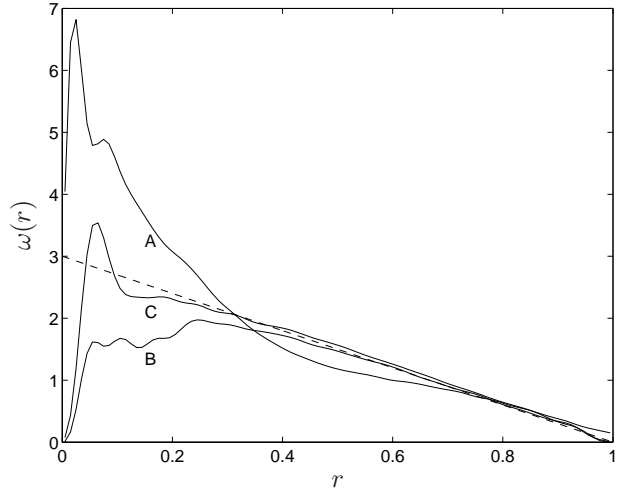}
\caption{\label{fig: bootstrapping}
Results from applying the boot strapping method in \Sref{estimate} to phase space trajectories from a DPD simulation. Curve A shows $\omega(r)$ after one iteration, starting with the initial guess $\omega(r) = 0$. Curves B and C show $\omega(r)$ after the second and fourth iteration, respectively. The dashed line gives the exact value of $\omega(r)$ as used in the DPD simulation.}
\end{figure}

As explained in the previous section, we wish to demonstrate that it is possible to obtain good estimates of $\omega(r)$ from \Eref{eq: final derived omega}, not only in the $\tau \rightarrow 0$ limit which is attainable in DPD but generally unavailable for a coarse grained system, but also for larger values of $\tau$. By using the boot strapping procedure outlined above for simulation data in the region $\tau \in [0.1, 0.25]$, we show in \Fref{fig: bootstrapping} the sequence of resulting $\omega(r)$ curves for the first boot strapping steps. 
Convergence towards the correct functional form of $\omega(r)$ is fast. Within the first four iterations, the method has approached the correct $\omega$-values used in the DPD simulation for most $r$-values. 
The conservative forces prevent the beads from coming arbitrarily close. This leads to a lack of data for small values of $r$, and explains the poor performance of the method in this region. For these values of $r$, we recommend to either set $\omega(r)$ to zero, or to use the value $\omega(r^*)$, where $r^*$ is the smallest value of $r$ with reliable statistics in the simulations.

\begin{figure} 
\centering 
\psfrag{t}[t][]{$t$}
\psfrag{-dDdpiDdpjoverdt}[][t]{$- \frac{\partial}{\partial \tau} \Expt{\delta_t (\delta_F \bp_i)\cdot\delta_t (\delta_F \bp_j)}_{r}$}

\includegraphics[width=8cm]{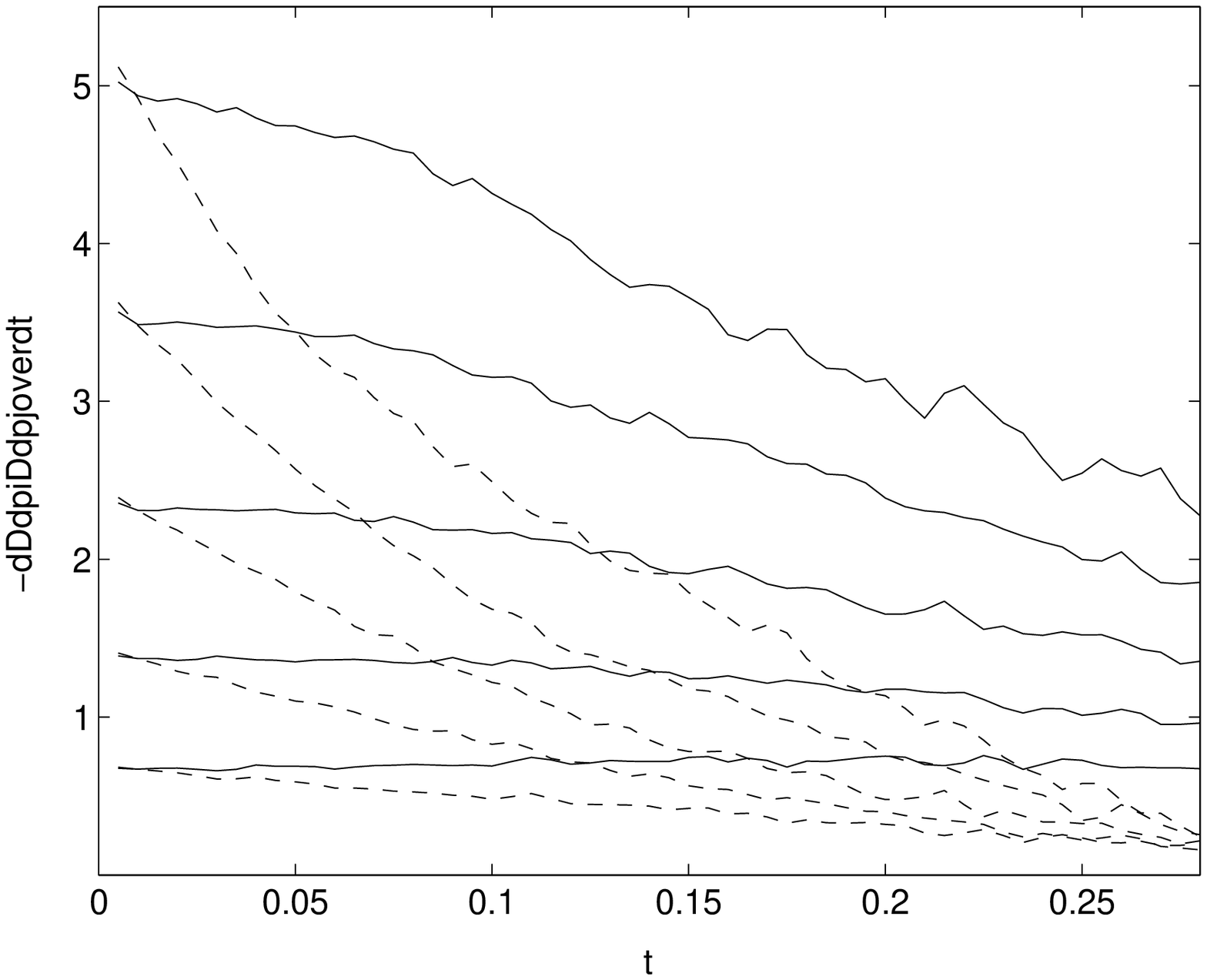}
\caption{\label{fig: force covariance removing correct omega}
The slope of the momentum change covariance as a function of $\tau$ for different values of $r$. The curves are calculated from phase space trajectories of two DPD simulations, using different definitions of $\delta_F \bF_i$. The solid and dashed lines correspond to using \Eref{eq: removing full deterministic force} and \Eref{eq: removing only conservative force}, respectively. The different curves in the figure correspond to the r-values 0.245 (top), 0.365, 0.485, 0.605 and 0.725 (bottom). The conservative force was set to $0$ in the simulations to demonstrate as clearly as possible the effect of not removing the dissipative force when calculating  $\delta_t(\delta_F \bp_i)$ . 
}
\end{figure}

To emphasize the importance of removing the full deterministic force (i.e. both conservative and dissipative forces) from the total force when calculating $\delta_t(\delta_F \bp_i)$, we show in \Fref{fig: force covariance removing correct omega} the slope of the momentum change covariance, as defined in \Eref{eq: final derived omega}, for two different cases. In the first case (solid lines), $\delta_F \bF_i$ is defined as 
\begin{equation}
\label{eq: removing full deterministic force}
\delta_F \bF_i = \bF_i - \bF_i^C - \bF_i^D,
\end{equation}
and in the second case (dashed lines),
\begin{equation}
\label{eq: removing only conservative force}
\delta_F \bF_i = \bF_i - \bF_i^C,
\end{equation}
where $\bF_i$ is the total force acting on particle $i$, $\bF_i^C$ is the conservative DPD force, and $\bF_i^D$ the dissipative DPD force. In the limit of $\tau \rightarrow 0$, both methods converge to the correct value of $\omega(r)$, as can be seen from \Fref{fig: force covariance removing correct omega}, but with $\delta_F \bF_i$ defined by \Eref{eq: removing full deterministic force}, there exists a plateau of small $\tau$-values for which the force covariance is approximately constant. Measuring $\omega^2(r)$ anywhere in this region gives approximately the correct value, and as shown previously, using $\tau$-values as far out as $\tau \in [0.1, 0.25]$ still reproduces $\omega(r)$ using the boot strapping procedure.

\section{Discussion}

We have discussed how to derive a dissipative particle dynamics from a detailed microscopic system, for example a molecular dynamics simulation. As a coarse grained model of a mechanical system, DPD has several advantages compared to many other models. Most importantly, by construction the DPD dynamics respects fundamental symmetries of the underlying dynamics; it is Galilean invariant, and therefore both linear and angular momentum are locally conserved by the interactions.

In this paper, we have established two important properties of our method:
First, that the framework is consistent; if the dynamics is on the DPD form, we can use the method to accurately reconstruct all terms in the equations of motion. Through our adaption of the Mori-Zwanzig projection operator methods, we argue that this provides a clear and quantifiable connection to the underlying degrees of freedom.

Second, the method successfully reconstructs the dynamics without using the shortest time-scale  of the particle trajectories. If the detailed model shows a strong time-scale separation  it is possible to use \Eref{eq:dpidpj} directly to estimate the effective stochastic interactions on the coarse grained level. In many cases however -- e.g. when DPD is used as a coarse-grained representation of a molecular system -- the time scale separation is not very significant. It is therefore essential that our method is able to reconstruct the dynamics without using the correlations of the system at the shortest time scale, which we demonstrate in \Fref{fig: minus_DdpiDpdj_FC25}. 

The price we pay for not using the short-time properties is that we cannot use a direct method to measure the shape of $\omega(r)$, but are forced to use an iterative scheme. This is because $\omega(r)$ is estimated from the stochastic force, and in order to extract this force from the dynamics we need to subtract influence of the deterministic (conservative and dissipative) forces from the particle trajectories, which then depend on the $\omega(r)$ that we try to estimate in the first place. However, we have found that starting from the initial $\omega(r) = 0$, and using the estimation as a fixed-point scheme, leads to rapid convergence (\Fref{fig: bootstrapping}); the likely cause for this is that the dissipative component of the deterministic force is typically dominated by the conservative component, so that already the first iteration leads to an $\omega(r)$ not far from the correct one.

The main advantages of our method is perhaps that it gives the appropriate magnitude of dissipative and stochastic forces for the coarse-grained system to be consistent with the underlying dynamics; hence, if the coarse grained dynamics is averaging, so that fluctuations are not important, the resulting $\omega(r) \approx 0$, and if the rapidly fluctuating degrees of freedom act as white noise on the coarse-grained dynamics, $\omega(r)$ captures this effect. This is in contrast to most thermostats used in MD, where in principle the thermostat is used to stabilize the dynamics and is not considered to be an integral part of the dynamics.

\ack
This work was funded (in part) by the Programmable Artificial Cell Evolution project (EU integrated project FP6-IST-FET PACE), by EMBIO (a European Project in the EU FP6 NEST Initiative), by the Research Council of Norway and the Research Council of Sweden. 

\section*{References}
\bibliography{article}

\end{document}